# Real-Space Double Crossing Reveals Symmetry-Driven Dispersion Back-Bending of Surface Phonon Polaritons on α-Quartz

*Ryoga Odawara*[1*‡], *Kotaro Shirahata*[1‡], *Aozora Ohi*[1‡], *Shun Hashiyada*[1,2,3*], *and Yukio Kawano*[1,4,5*]

[1]Department of Electrical, Electronic, and Communication Engineering, Faculty of Science and Engineering, Chuo University, 1-13-27 Kasuga, Bunkyo-ku, Tokyo 112-8551, Japan

[2]Research Institute for Electronic Science, Hokkaido University, Kita 21, Nishi 10, Sapporo, Hokkaido 001-0021, Japan

[3]PRESTO Individual Researcher, Japan Science and Technology Agency, 4-1-8 Honcho, Kawaguchi-shi, Saitama 332-0012, Japan

[4]National Institute of Informatics, 2-1-2 Hitotsubashi, Chiyoda-ku, Tokyo 101-8430, Japan

[5]Kanagawa Institute of Industrial Science and Technology, 705-1 Imaizumi, Ebina-shi, Kanagawa 243-0435, Japan

‡Contributed equally to this work

**KEYWORDS**

surface phonon polaritons, dispersion back-bending, anisotropic polar crystals, α-quartz, s-SNOM, nano-FTIR spectroscopy

ABSTRACT

Surface phonon polaritons (SPhPs) in anisotropic crystals exhibit direction-dependent propagation and dispersion, yet how their dispersion back-bending manifests in real space has not been quantitatively resolved. Here, using nano-Fourier-transform infrared spectroscopy based on scattering-type scanning near-field optical microscopy, we image SPhP fringes on an α-quartz surface for propagation parallel and perpendicular to the optic axis. Near 1160 $cm^{-1}$, the fringe periods for the two directions cross twice, providing a real-space signature of back-bending. Spatial Fourier analysis attributes the dominant fringe to edge-launched SPhPs and enables reconstruction of the intrinsic dispersion. Inverse Fourier analysis further visualizes the spatial attenuation and continuous evolution of the fringe wavelength near the resonance. The reconstructed dispersions reveal that the polarization-selective E-mode phonon produces direction-dependent back-bending and, through its damping, reduces the visibility of the back-bending branch. These results establish how symmetry-controlled dispersion manifests in real-space surface polariton propagation on a bulk anisotropic crystal.

Surface phonon polaritons (SPhPs) are hybrid quasiparticles formed by the coupling between photons and optical phonons at the surfaces of polar crystals.[1,2] They combine subwavelength optical confinement with comparatively low propagation losses in the mid-infrared (MIR) and the terahertz spectral ranges,[2,3] providing a promising platform for superlensing,[4,5] nanocavities,[6] and surface-enhanced spectroscopy.[7,8] In anisotropic crystals, the tensorial dielectric response further gives rise to direction-dependent phonon-polariton dispersions with elliptical or hyperbolic isofrequency contours.[8–13] Such anisotropy enables unusual propagation phenomena, including negative phase velocity,[11] sub-diffractional focusing,[12] and highly directional transport.[13] More recently, ghost hyperbolic polaritons have been demonstrated in bulk calcite, where tilting the optic axis (OA) relative to the surface produces oblique wavefronts with simultaneous phase oscillation and exponential attenuation.[14] These advances highlight crystal symmetry and OA orientation as key degrees of freedom for controlling polariton propagation.

The SPhP dispersion of α-quartz, a canonical anisotropic polar crystal, and its characteristic back-bending were established by pioneering attenuated total reflection measurements.[15–17] However, such momentum-space measurements are inherently blind to how the back-bending manifests in real-space propagation. Laser-induced periodic surface structures later provided indirect evidence of SPhP propagation,[18] while scattering-type scanning near-field optical microscopy (s-SNOM) studies probed local SPhP resonances on α-quartz.[19,20] Nevertheless, how the broadband polariton wavelength, attenuation, and branch visibility across the back-bending region has not been resolved quantitatively in real space. Bridging this gap between complex-valued momentum-space dispersion and real-space propagation has become particularly pertinent given recent near-field studies of polaritons in low-symmetry bulk crystals, including calcite,[14,21,22] $\beta$-$Ga_2O_3$,[23,24] $CdWO_4$,[25] $Y_2SiO_5$,[26] and $YVO_4$.[27] α-Quartz, whose momentum-space response is well established, offers an ideal benchmark for this purpose.

Here, we use broadband nano-Fourier-transform infrared (nano-FTIR) spectroscopy based on s-SNOM to image anisotropic SPhP propagation on an X-cut α-quartz surface and quantitatively reconstruct its dispersion. Hyperspectral near-field line profiles reveal that the relative fringe periods for propagation parallel and perpendicular to the OA cross twice (double crossing) within an 8 $cm^{-1}$ interval near 1160 $cm^{-1}$, providing a direct real-space signature of dispersion back-bending that is visible even in the raw line profiles. Spatial Fourier analysis reconstructs the intrinsic wavevector–frequency dispersion across the Reststrahlen band, while inverse Fourier analysis visualizes the accompanying wavelength evolution and enhanced spatial attenuation. The results reveal how the polarization-selective E-mode phonon produces direction-dependent back-bending and, through its damping-induced dielectric loss, reduces the visibility of the back-beding branch. Our study thus reveals how symmetry- and loss-controlled dispersion back-bending manifests in real space on a bulk anisotropic polar crystal.

We first characterize the anisotropic optical response of α-quartz that governs its SPhP dispersion. Because α-quartz is a uniaxial polar crystal, its dielectric tensor can be written in the principal-axis frame as $\hat{\varepsilon} = \mathrm{diag}(\varepsilon_\perp, \varepsilon_\perp, \varepsilon_\parallel)$, where $\varepsilon_\parallel$ and $\varepsilon_\perp$ denote the components parallel and perpendicular to the OA ([0001]), respectively. Figure 1a schematically illustrates the polarized specular-reflectance measurement performed using FTIR spectroscopy to characterize this optical anisotropy. We used an X-cut α-quartz substrate, in which the OA lies in the surface plane and the surface normal is along $[11\bar{2}0]$. Figure 1b shows the reflectance spectra measured with the incident electric field parallel ($\boldsymbol{E} \parallel$ OA, red curve) and perpendicular ($\boldsymbol{E} \perp$ OA, blue curve) to the OA. In both polarization configurations, α-quartz exhibits a broad high-reflectivity Reststrahlen band (approximately 1070 to 1240 $cm^{-1}$), within which the corresponding permittivity component becomes negative, allowing the surface to support SPhPs. A pronounced dip appears near 1160 $cm^{-1}$ only for $\boldsymbol{E} \perp$ OA. This feature, reported previously,[28,29] arises from an infrared-active E mode phonon near 1160 $cm^{-1}$, no infrared-active $A_2$ mode exists in this range.[29,30] Because E modes are infrared-active only for electric fields perpendicular to the OA, the dip appears exclusively for $\boldsymbol{E} \perp$ OA.

For quantitative analysis, the complex dielectric functions, $\varepsilon(\omega) = \varepsilon'(\omega) + i\varepsilon''(\omega)$, were obtained from the reflectance spectra by a Kramers–Kronig (K–K) analysis (Figures 1c, d). The resulting dielectric functions reveal anisotropic shifts in the transverse-optical (TO, $\omega_{\mathrm{TO}}$) and longitudinal-optical (LO, $\omega_{\mathrm{LO}}$) phonon frequencies. From the lower and upper sign change of $\mathrm{Re}(\varepsilon) = \varepsilon'$, which delimit the negative-permittivity bands, we estimated the effective TO and LO phonon frequencies, respectively. The resulting values are $\omega_{\mathrm{TO},\parallel} \approx 1073\ \mathrm{cm}^{-1}$, $\omega_{\mathrm{LO},\parallel} \approx 1241\ \mathrm{cm}^{-1}$, $\omega_{\mathrm{TO},\perp} \approx 1065\ \mathrm{cm}^{-1}$, and $\omega_{\mathrm{LO},\perp} \approx 1229\ \mathrm{cm}^{-1}$. These frequencies and the overall anisotropic response agree to within approximately 10 $cm^{-1}$ with previously reported values.[29,31] The small frequency offsets may arise from truncation errors in the K–K transform; as discussed below for Figure 5, they do not affect the reconstructed dispersion. The anisotropy is particularly pronounced in the imaginary part: $\mathrm{Im}(\varepsilon_{\perp}) = \varepsilon''_{\perp}$ exhibits a strong peak near 1160 $cm^{-1}$, whereas $\varepsilon''_{\parallel}$ remains comparatively small. The peak position agrees with the reported E-mode phonon frequency.[28–31] This component-selective resonance and its associated loss provide the physical basis for the direction-dependent SPhP propagation discussed below.

We next investigate the real-space propagation of the anisotropic SPhPs using nano-FTIR spectroscopy. As illustrated in Figure 2a, broadband p-polarized MIR light ($E_{\mathrm{in}}$) is focused by a parabolic mirror onto the s-SNOM tip and an Au disk fabricated on the α-quartz surface at a polar angle of incidence $\theta$. The illuminated disk edge launches SPhPs ($E_{\mathrm{el}}$),[32] whereas the tip launches SPhPs ($E_{\mathrm{tl}}$) that propagate to the disk edge, are reflected, and return to the tip.[33] At the tip position, these SPhP fields interfere with the directly backscattered field ($E_{\mathrm{bs}}$). Consequently, the observed fringe periods ($\Lambda$) differ from the intrinsic SPhP wavelength ($\lambda_{\mathrm{SPhP}}$).[33–36] Neglecting terms quadratic in the SPhP fields, we approximated the detected signal as

$$I(x) \approx |E_{\mathrm{bs}}|^2 + 2 \sum_{j=\mathrm{tl,el}} |E_{\mathrm{bs}}||E_j(x)| \cos[\phi_{\mathrm{bs}}(x) - \phi_j(x)]. \tag{1}$$

Here, $x$ is the coordinate along the scan direction, and $\phi_{\mathrm{bs}}$ and $\phi_j$ are the phases of the backscattered field and the corresponding SPhP field, respectively. For the tip-launched mode, the SPhP travels a round trip between the tip and edge, so that the fringe period is[33,35]

$$\Lambda_{\mathrm{tl}} = \frac{\lambda_{\mathrm{SPhP}}}{2}. \tag{2}$$

By contrast, the fringe period of the edge-launched mode ($\Lambda_{\mathrm{el}}$) is determined by geometric phase matching:

$$\Lambda_{\mathrm{el}}(\theta, \varphi) = \frac{\lambda_0}{-\sin\theta \sin\varphi + \sqrt{\sin^2\theta\,(\sin^2\varphi - 1) + n^2}}. \tag{3}$$

Here, $n = \lambda_0/\lambda_{\mathrm{SPhP}}$ is the effective SPhP index, $\lambda_0$ is the free-space wavelength, and $\varphi$ is the azimuthal angle of the incident wavevector relative to the disk edge (for a detailed derivation, see refs 33–35). Figure 2b illustrates the two measurement configurations. The scan direction $x$ was oriented either parallel ($x \parallel$ OA) or perpendicular ($x \perp$ OA) to the OA while remaining antiparallel to the in-plane projection of the incident wavevector. In the present geometry, $\varphi = 270°$, and eq 3 therefore reduces to

$$\Lambda_{\mathrm{el}}(\theta, 270°) = \frac{\lambda_0}{\sin\theta + n}. \tag{4}$$

Figures 3a and 3b show hyperspectral near-field amplitude images obtained for propagation parallel and perpendicular to the OA, respectively. The interference fringes emerge above approximately 1090 $cm^{-1}$ in the parallel configuration and above approximately 1070 $cm^{-1}$ in the perpendicular configuration. This difference in onset frequency is consistent with the anisotropic TO-phonon frequencies ($\omega_{\mathrm{TO}}$) determined from Figure 1. We then focus on the spectral region near 1160 $cm^{-1}$, where the dielectric anisotropy is most pronounced. Figure 3c shows representative near-field line profiles together with fits using a damped-sinusoidal model (yellow solid lines). At 1159 $cm^{-1}$, the fitted fringe periods differ clearly, $\Lambda_{\parallel} \approx 4.07$ μm and $\Lambda_{\perp} \approx 4.32$ μm, even though they are measured at the same frequency. Remarkably, the ordering of the two fringe periods reverses twice within an 8 $cm^{-1}$ interval: $\Lambda_{\parallel} > \Lambda_{\perp}$ at 1155 $cm^{-1}$, $\Lambda_{\parallel} < \Lambda_{\perp}$ at 1159 $cm^{-1}$, and $\Lambda_{\parallel} > \Lambda_{\perp}$ again at 1163 $cm^{-1}$. This double crossing is directly visible in the line profiles and provides a real-space signature of dispersion back-bending without requiring Fourier-space reconstruction.

To determine the SPhP wavevectors, we performed a fast Fourier transform (FFT) of the near-field line profiles in Figures 3a and 3b along the $x$ direction. Following established procedures,[8,11,33] we obtained the FFT-magnitude maps shown in Figure 4a; details of the data processing are provided in Section 1 of the Supporting Information. The solid and dashed curves indicate the calculated fringe wavevectors of the edge-launched mode, $k_{\mathrm{el}} = 2\pi/\Lambda_{\mathrm{el}}$, and the tip-launched mode, $k_{\mathrm{tl}} = 2\pi/\Lambda_{\mathrm{tl}}$, respectively. Both are determined by the intrinsic complex SPhP wavevector (eqs 2 and 4), which for propagation along a principal axis, is given by[1,15,16,32]

$$k_{\mathrm{SPhP}} = k'_{\mathrm{SPhP}} + ik''_{\mathrm{SPhP}} = \frac{\omega}{c}\sqrt{\frac{\varepsilon_x\varepsilon_z - \varepsilon_z}{\varepsilon_x\varepsilon_z - 1}}. \tag{5}$$

Here, $\varepsilon_x$ and $\varepsilon_z$ are the components of the dielectric tensor along the SPhP propagation direction and the surface normal, respectively. Because the surface normal of the X-cut substrate is perpendicular to the OA, $\varepsilon_z = \varepsilon_\perp$ in both measurement configurations, whereas $\varepsilon_x$ depends on the propagation direction. Substituting the relevant tensor components into eq 5 gives

$$k_{\mathrm{SPhP}}^{\parallel\,\mathrm{OA}} = \frac{\omega}{c}\sqrt{\frac{\varepsilon_\parallel \varepsilon_\perp - \varepsilon_\perp}{\varepsilon_\parallel \varepsilon_\perp - 1}}, \qquad k_{\mathrm{SPhP}}^{\perp\,\mathrm{OA}} = \frac{\omega}{c}\sqrt{\frac{\varepsilon_\perp}{\varepsilon_\perp + 1}}, \tag{6}$$

for propagation parallel and perpendicular to the OA, respectively. The real part, $k'_{\mathrm{SPhP}}$, determines the SPhP wavelength, whereas the imaginary part, $k''_{\mathrm{SPhP}}$, describes the spatial attenuation of the propagating field[32].

The dielectric functions obtained from the K-K analysis were used to calculate $k_{\mathrm{SPhP}}$, from which we derived $\lambda_{\mathrm{SPhP}} (= 2\pi/k'_{\mathrm{SPhP}})$, $k_{\mathrm{tl}}$, and $k_{\mathrm{el}}$ (eqs 2 and 4). The effective polar angle of incidence depends on the alignment of the parabolic mirror and may therefore deviate[36] from the nominal instrument angle of 60°. We treated $\theta$ as a single global geometric parameter, common to all frequencies and for both configurations, and determined $\theta = 45°$ from the best overall agreement between the measured FFT ridges and the calculated edge-launched dispersion. With this single effective angle, the calculated $k_{\mathrm{el}}$ curves follow the dominant FFT ridges in both configurations, which is consistent with a predominantly edge-launched origin of the observed fringes.

Clear back-bending is observed near 1160 cm$^{-1}$ in both the calculated $k_{\mathrm{el}}$ curves and the experimental FFT maps. The back-bending is more pronounced for propagation perpendicular to the OA. This difference follows from the dielectric-tensor components entering the surface-mode dispersion. For propagation parallel to the OA, $(\varepsilon_x, \varepsilon_z) = (\varepsilon_\parallel, \varepsilon_\perp)$, whereas for propagation perpendicular to the OA, $(\varepsilon_x, \varepsilon_z) = (\varepsilon_\perp, \varepsilon_\perp)$. Because the polarization-selective E-mode resonance appears only in $\varepsilon_\perp$ by symmetry, it affects only the out-of-plane component ($\varepsilon_z$) in the parallel configuration, but both relevant components ($\varepsilon_x, \varepsilon_z$) in the perpendicular configuration. The resonance therefore produces stronger dielectric loss and a more pronounced modification of the dispersion in the perpendicular configuration. In the ideal lossless limit, the surface-mode wavevector would diverge as the corresponding resonance condition is approached: $\mathrm{Re}(\varepsilon_\parallel \varepsilon_\perp) \approx 1$ for propagation parallel to the OA and $\mathrm{Re}(\varepsilon_\perp) \approx -1$ for propagation perpendicular to the OA[37]. The finite damping of the E-mode phonon regularizes this divergence, limiting the wavevector to a finite maximum and producing the observed back-bending.

Figures 4b and 4c show the amplitude ($|E_{\mathrm{el}}|$) and phase ($\phi_{\mathrm{el}}$) maps of the edge-launched SPhPs reconstructed by inverse FFT (iFFT). To isolate the edge-launched contribution and suppress truncation artifacts, we applied a Hann window centered on the calculated $k_{\mathrm{el}}$ curves before performing the iFFT (see Section 2 of the Supporting Information). The normalized amplitude maps in Figure 4b reveal rapid spatial decay between approximately 1150 and 1160 $\mathrm{cm}^{-1}$. This decay is attributed to the enhanced dielectric loss associated with the E-mode phonon and is substantially stronger in the perpendicular configuration. The white curves indicate the calculated propagation length, $L_{\mathrm{SPhP}} = 1/\mathrm{Im}(k_{\mathrm{SPhP}})$ .[32] The measured spatial decay is qualitatively consistent with the calculated reduction in $L_{\mathrm{SPhP}}$. The phase maps in Figure 4c further reveal the continuous evolution of the edge-launched fringe period, $\Lambda_{\mathrm{el}}$. Near 1160 $\mathrm{cm}^{-1}$, $\Lambda_{\mathrm{el}}$, which otherwise decreases with increasing frequency, temporarily increases and then decreases again above the resonance. This continuous wavelength evolution is a real-space manifestation of the back-bending observed in the FFT dispersion maps.

To quantitatively reconstruct the intrinsic SPhP dispersion, we determined the experimental edge-launched wavevectors, $k_{\mathrm{el}}^{\mathrm{exp}}$, by fitting Lorentzian functions to the step-windowed FFT-magnitude profiles at each frequency (see Section 3 of the Supporting Information). The resulting values were converted to the intrinsic SPhP wavevectors using

$$k'_{\mathrm{SPhP}}(\theta) = k_{\mathrm{el}}^{\mathrm{exp}} - k_0 \sin\theta\,, \tag{7}$$

where $k_0 = 2\pi/\lambda_0$. The reconstructed wavevectors for propagation parallel ($k'^{\parallel\mathrm{OA}}_{\mathrm{SPhP}}$) and perpendicular ($k'^{\perp\mathrm{OA}}_{\mathrm{SPhP}}$) to the OA are plotted as red circles in Figure 5a and blue squares in Figure 5b, respectively. The background color maps show the imaginary part of the p-polarized Fresnel reflection coefficient, $\mathrm{Im}(r_{\mathrm{pp}})$, calculated from the K–K-derived dielectric functions using a transfer-matrix method.[38] The solid curves show SPhP dispersions calculated using the finite-element method (FEM) with the oscillator-model dielectric functions determined by ellipsometry in ref 31 (Table S1). Although the transfer-matrix and FEM calculations are based on independently obtained dielectric data, they reproduce the same overall dispersion and back-bending features. This agreement indicates that the small frequency offsets in the K–K-derived dielectric functions do not affect the principal features of the dispersion analysis.

Overall, the reconstructed experimental dispersions follow the calculated anisotropic SPhP branches and reproduce the back-bending near 1160 $\mathrm{cm}^{-1}$ in both propagation configurations. The different wavevectors along the two principal in-plane directions are consistent with an anisotropic, approximately elliptical isofrequency contour of the bound surface-polariton branch.

Near the E-mode resonance, the experimentally extracted wavevectors do not fully trace the smooth, continuous branches obtained from the FEM calculations. In the numerical calculation, the complex-wavevector SPhP solution can be followed continuously even when the imaginary part of the wavevector becomes large. In the experiment, however, a large $\mathrm{Im}(k_{\mathrm{SPhP}})$ strongly suppresses the near-field signal of propagating SPhPs. Consequently, the real-space fringes decay over a short distance, and the corresponding FFT peaks become weak and broadened, making their center positions difficult to determine. This loss of experimental visibility is more severe in the perpendicular configuration because the E-mode resonance contributes to both relevant dielectric-tensor components. The seemingly continuous trajectory in the parallel configuration and the apparent splitting in the perpendicular configuration may therefore reflect differences in the experimental visibility of the strongly damped branch rather than from a fundamental difference in the underlying dispersion.

Despite the reduced visibility near the resonance, the reconstructed wavevectors exhibit local maxima at approximately 1159 $cm^{-1}$ for propagation parallel to the OA and 1157 $cm^{-1}$ for propagation perpendicular to the OA. Falge and Otto[15] reported that, for these propagation geometries, the SPhP resonance frequencies lie jus below the E-mode phonon frequency ($\omega_{\mathrm{E-mode}}$), at $\omega_{\mathrm{SP}}^{\parallel\mathrm{OA}} = 1159\ \mathrm{cm}^{-1}, \omega_{\mathrm{SP}}^{\perp\mathrm{OA}} = 1157\ \mathrm{cm}^{-1}$, in agreement with the present values to within the spectral resolution. This agreement supports the assignment of the observed back-bending to the E-mode phonon. It further indicates that the direction dependence of the back-bending originates from the symmetry-controlled dielectric response of the crystal.

In conclusion, we have revealed how anisotropic SPhP dispersion back-bending manifests in real space on an X-cut α-quartz surface. The interference fringes for propagation parallel and perpendicular to the OA exhibit a double crossing of their relative periods near 1160 $cm^{-1}$, providing a direct real-space signature of back-bending without requiring Fourier analysis. Spatial FFT analysis attributes the dominant fringe to edge-launched SPhPs and enables quantitative reconstruction of the intrinsic wavevector–frequency dispersion across the Reststrahlen band. The corresponding iFFT analysis visualizes the continuous wavelength evolution and enhanced spatial attenuation near the E-mode resonance. The reconstructed dispersions show that the extent and experimental visibility of the back-bending depend strongly on the propagation direction, revealing how the polarization-selective E-mode phonon, anisotropic dielectric response, and phonon damping govern the observed polariton propagation. These results how symmetry- and loss-controlled complex-valued dispersion manifests in real space on a bulk anisotropic polar crystal. More broadly, they provide a basis for controlling surface-polariton transport through crystal orientation and may facilitate exploration of ghost polaritons[14] in α-quartz with the OA tilted relative to the surface.

**Methods**. *Sample fabrication*. For the α-quartz substrates, we used X-cut single-crystal substrates (500 μm thick) with the OA (c axis, [0001]) parallel to the surface. After the substrates were cleaned with organic solvents, a positive electron-beam resist (ZEP520A-7, ZEON) was spin-coated and patterned by electron-beam lithography system (F7000-VD02, ADVANTEST). Subsequently, a Ti adhesion layer (thickness 5 nm) and an Au film (60 nm) were then deposited by electron-beam vacuum deposition system. Finally, lift-off yielded an Au disk (10 μm in diameter) on the α-quartz surface.

*s-SNOM measurements*. Line-scan measurements of SPhPs were performed using a scattering-type near-field optical microscope (neaSNOM, attocube systems) equipped with a nano-FTIR module. A broadband MIR laser source based on an optical parametric oscillator and difference frequency generation system (FemtoFiber pro laser system, TOPTICA) served as the light source. Broadband infrared p-polarized light was focused onto the tip of a Pt/Ir-coated atomic force microscope (AFM) probe (ATEC-NcPt, NANOSENSORS) using a parabolic mirror. The AFM probe was operated in tapping mode at a frequency of $\Omega \approx 253$ kHz with an amplitude of ~110 nm. The light scattered by the tip was analyzed with an asymmetric Michelson interferometer in the nano-FTIR module and detected with a mercury telluride (MCT) detector via. To suppress the far-field background, the detection signal was demodulated at the second harmonic ($2\Omega$) of the tapping frequency. In line-scan measurements, an interferogram was acquired for each pixel, and the near-field amplitude and phase at each position was obtained by applying a Fourier transform.

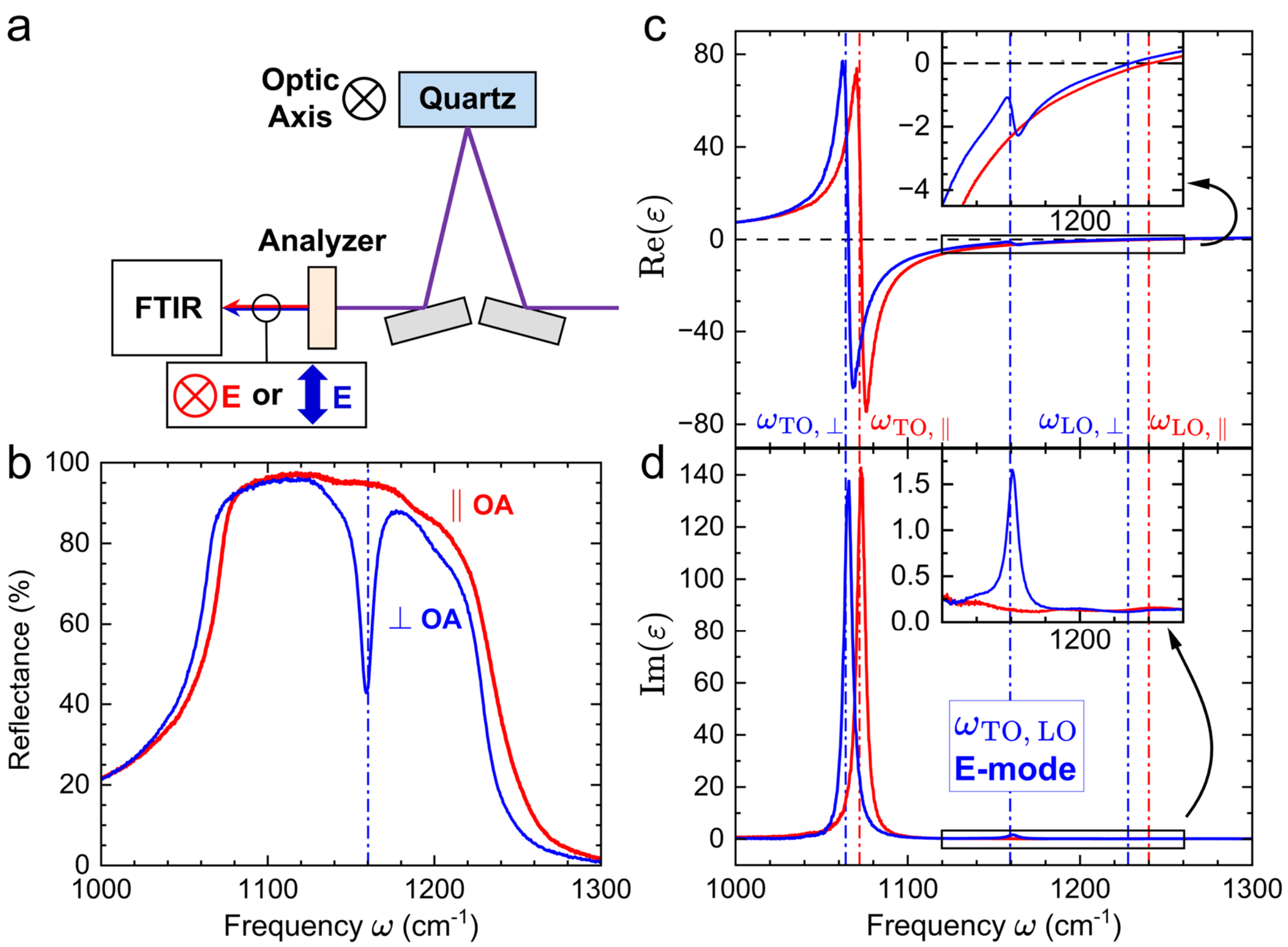

**Figure 1.** Anisotropic Reststrahlen band in α-quartz. (a) Schematic of the polarized specular-reflectance measurement for incident light polarized parallel (red) and perpendicular (blue) to the optic axis (OA). (b) Measured infrared reflectance spectra. (c) Real and (d) imaginary complex dielectric permittivity, $\varepsilon = \varepsilon' + i\varepsilon''$, extracted by Kramers–Kronig (K–K) analysis. Dash-dotted lines indicate the TO and LO phonon frequencies ($\omega_{\mathrm{TO}}$ and $\omega_{\mathrm{LO}}$) for each polarization, and in (b) and (d), the E-mode phonon frequency.

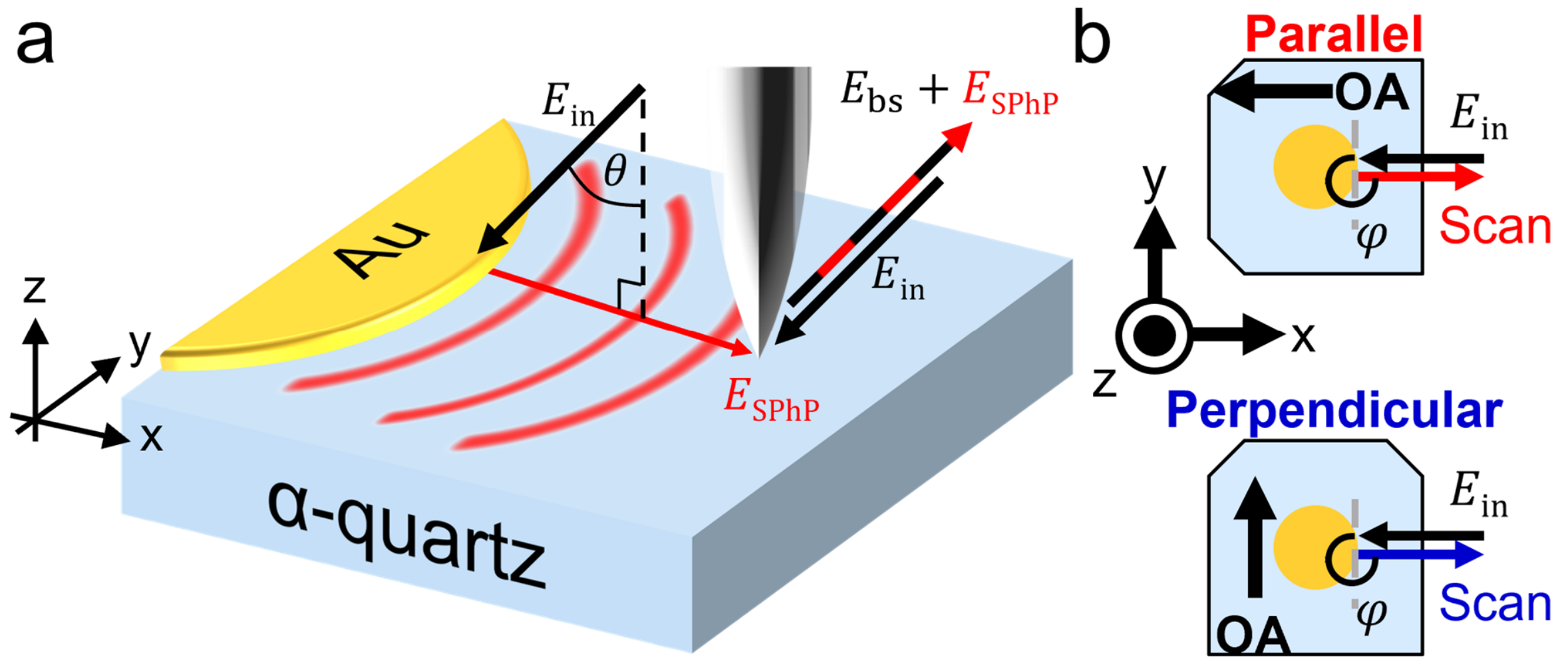

**Figure 2.** Experimental setup for real-space nanoimaging of SPhPs with an s-SNOM-based nano-FTIR system. (a) Schematic of the excitation and detection path of SPhPs launched at the Au disk edge. Broadband MIR p-polarized light ($E_{\rm in}$) illuminates both the Au disk edge and the s-SNOM tip. The field scattered by the tip results from interference between the SPhP fields ($E_{\rm SPhP} = E_{\rm el} + E_{\rm tl}$) and directly back-scattered light ($E_{\rm bs}$). The total field is detected as the s-SNOM signal. The s-SNOM tip is scanned away from the disk edge, along the edge normal, anti-parallel to the in-plane projection of the incident wavevector, to acquire hyperspectral line profiles of SPhPs. (b) Measurement configurations relative to the OA. Top: parallel configuration ($x \parallel$ OA); bottom: perpendicular configuration ($x \perp$ OA). The azimuthal angle of incidence relative to the disk edge is $\varphi = 270°$ in this configuration.

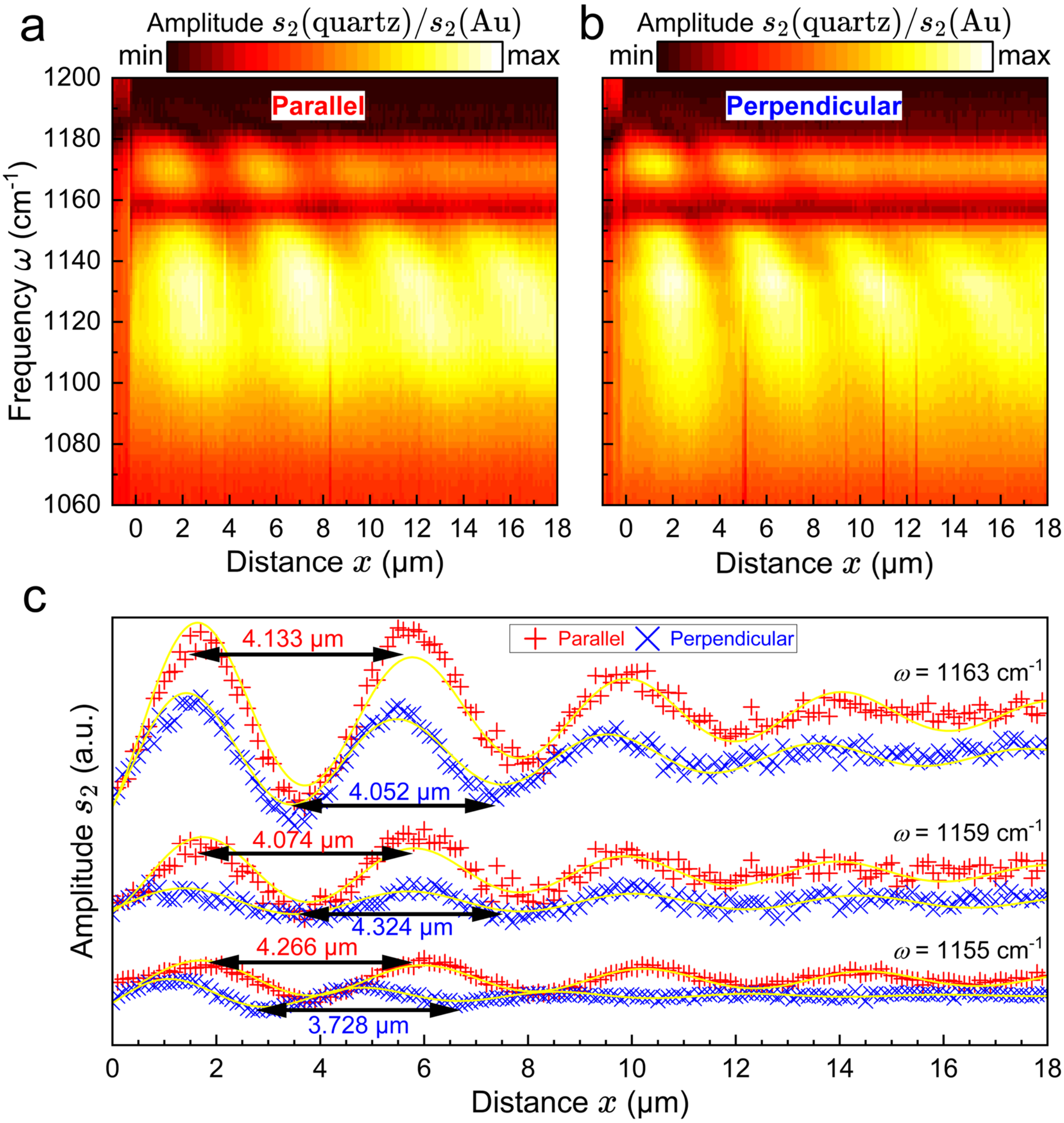

**Figure 3.** Real-space profiles of in-plane anisotropic SPhPs on α-quartz. Hyperspectral near-field amplitude line-scan images in the (a) parallel ($x \parallel$ OA) and (b) perpendicular ($x \perp$ OA) configurations. The second-harmonic amplitude $s_2$ is normalized to that at the center of the Au disk. The Au disk edge is located at $x = 0$. Owing to the optical anisotropy of the crystal, the interference fringe periods differ between the two configurations at the same frequency. (c)

Representative near-field amplitude line profiles (symbols) extracted from (a) and (b) at 1155, 1159, and 1163 $cm^{-1}$, with fits to a damped sinusoidal model (yellow lines). Profiles are vertically offset for clarity. Arrows indicate the fitted fringe period $\Lambda_{\parallel}$ (red) and $\Lambda_{\perp}$ (blue).

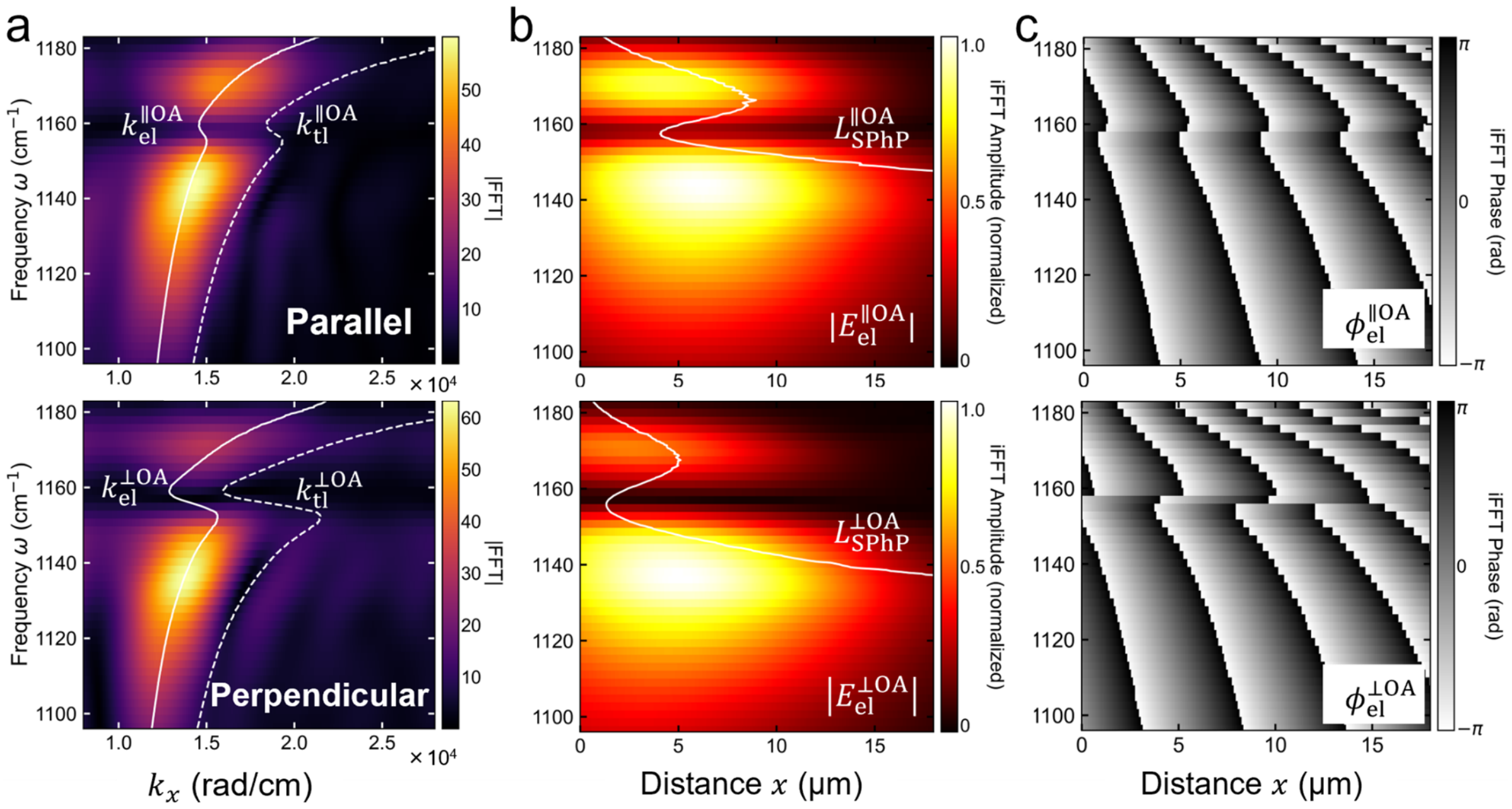

**Figure 4.** FFT-magnitude maps and reconstructed real-space maps of the edge-launched SPhPs for the parallel configuration (top) and perpendicular configuration (bottom). (a) FFT-magnitude maps of the line-scan images in Figure 3a, b. Solid and dashed curves show the calculated wavevectors of the edge-launched ( $k_{\mathrm{el}}$ ) and tip-launched ( $k_{\mathrm{tl}}$ ) SPhPs, respectively. (b, c) Reconstructed amplitude (b) and phase (c) maps of the edge-launched SPhPs. These maps were obtained by inverse FFT (iFFT) of the FFT maps after applying a Hann window centered on the calculated $k_{\mathrm{el}}$ curves. White curves in (b) indicate the calculated SPhP propagation length, $L_{\mathrm{SPhP}} = 1/\mathrm{Im}(k_{\mathrm{SPhP}})$.

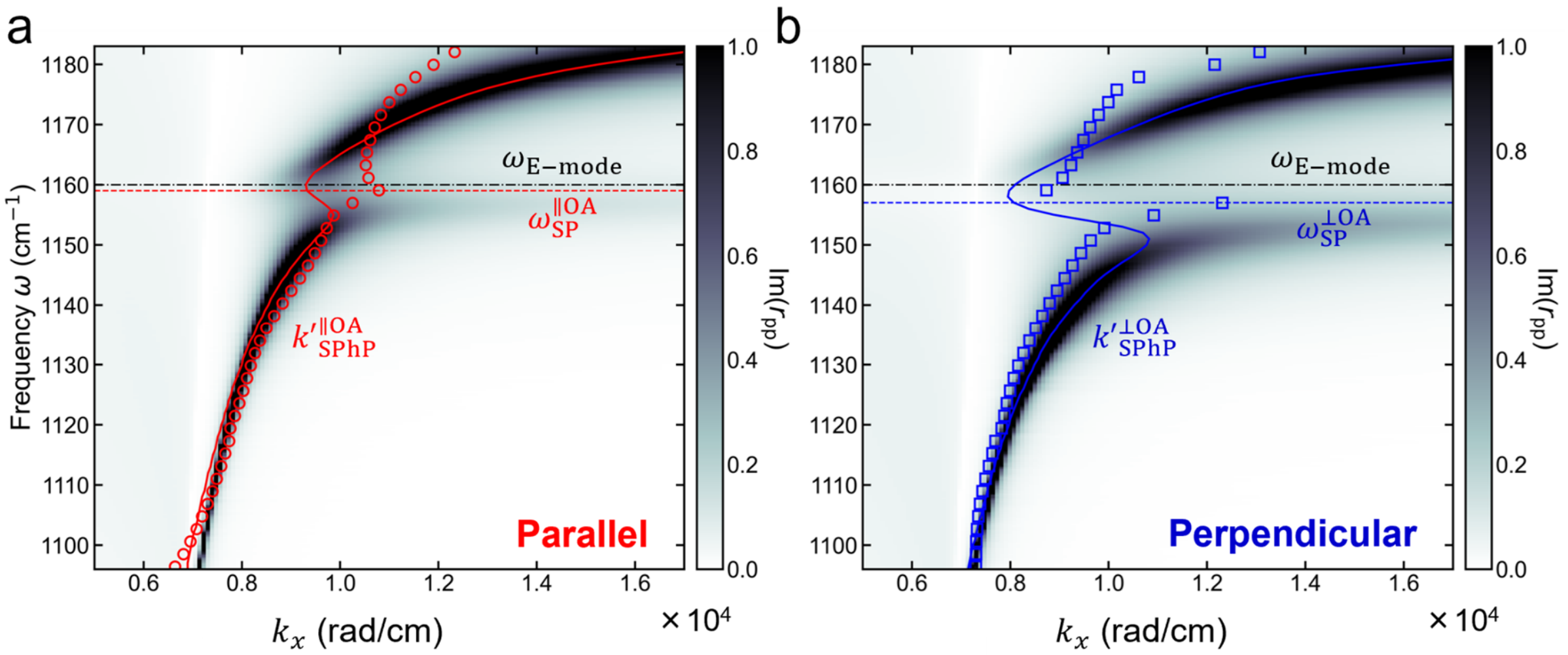

**Figure 5.** Reconstructed anisotropic SPhP dispersion on α-quartz. (a, b) Imaginary part of the p-polarized Fresnel reflection coefficient, $\text{Im}(r_{\text{pp}})$, calculated by the transfer matrix methods for the (a) parallel ($x \parallel$ OA) and (b) perpendicular configuration ($x \perp$ OA). Symbols: experimental $k'_{\text{SPhP}}$ (eq 7); solid lines: simulated $k'_{\text{SPhP}}$. Horizontal dash-dotted black lines correspond to the E-mode phonon frequency. Dashed red and blue lines mark the frequencies at which the experimental $k_{\text{SPhP}}$ reaches a local maximum, taken as the SPhP resonance frequency $\omega_{\text{SP}}$.

ASSOCIATED CONTENT

**Supporting Information.**

Data processing for fast Fourier transform (FFT) analysis of nano-FTIR line profiles. Reconstruction of edge-launched SPhPs using inverse FFT. Lorentzian fitting of step-windowed FFT-Magnitude profiles. Numerical simulations by finite-elements method. (PDF)

AUTHOR INFORMATION

**Corresponding Authors**

*Ryoga Odawara

E-mail: a20.b5bw@g.chuo-u.ac.jp

ORCID: 0009-0008-0815-7148

*Shun Hashiyada

E-mail: shun.hashiyada@es.hokudai.ac.jp

ORCID: 0000-0002-3229-538X

*Yukio Kawano

E-mail: kawano@elect.chuo-u.ac.jp

ORCID: 0000-0002-8456-2937

**Author Contributions**

S.H. and Y.K. conceived the project. K.S. and A.O. prepared the sample. R.O., K.S., and A.O. performed the experiments. R.O., K.S., A.O., and S.H. analyzed the data. S.H. and Y.K. supervised the project. R.O. wrote the original draft. R.O., K.S., A.O., S.H., and Y.K. reviewed and edited the manuscript. All authors have given approval to the final version of the manuscript. R.O., K.S. and A.O. contributed equally to this work.

**Notes**

The authors declare no competing financial interest.

ACKNOWLEDGMENT

This work was supported by the Japan Science and Technology Agency (JST) Mirai Program (Grant No. JPMJMI23G1), PRESTO (Grant No. JPMJPR25J6), and SPRING (Grant No. JPMJSP2170), Japan Society for the Promotion of Science (JSPS) KAKENHI (Grant Nos. JP21K14594, JP23H00169, JP23K04669, JP24K01288, JP25K01289, JP25H02154), Kanagawa Institute of Industrial Science and Technology, “Advanced Research Infrastructure for Materials and Nanotechnology in Japan (ARIM)” of the Ministry of Education, Culture, Sports, Science and Technology (MEXT) (Grant No. JPMXP1222UT1186), and the fabrication was conducted in Takeda Cleanroom, Center of The University of Tokyo for The Advanced Research Infrastructure for Materials and Data Hub.

During the preparation of this work, the authors used Claude Opus 5.5 (Anthropic) to improve the English grammar and readability of the manuscript and Gemini flash 3.8 (Google) to generate a draft of the Table Contents graphic, which was subsequently edited by the authors. The authors reviewed and edited all content as necessary and take full responsibility for the published article.

# *Supporting Information for* Real-Space Double Crossing Reveals Symmetry-Driven Dispersion Back-Bending of Surface Phonon Polaritons on α-Quartz

*Ryoga Odawara[1*‡], Kotaro Shirahata[1‡], Aozora Ohi[1‡], Shun Hashiyada[1,2,3*], and Yukio Kawano[1,4,5*]*

[1]Department of Electrical, Electronic, and Communication Engineering, Faculty of Science and Engineering, Chuo University, 1-13-27 Kasuga, Bunkyo-ku, Tokyo 112-8551, Japan

[2]Research Institute for Electronic Science, Hokkaido University, Kita 21, Nishi 10, Sapporo, Hokkaido 001-0021, Japan

[3]PRESTO Individual Researcher, Japan Science and Technology Agency, 4-1-8 Honcho, Kawaguchi-shi, Saitama 332-0012, Japan

[4]National Institute of Informatics, 2-1-2 Hitotsubashi, Chiyoda-ku, Tokyo 101-8430, Japan

[5]Kanagawa Institute of Industrial Science and Technology, 705-1 Imaizumi, Ebina-shi, Kanagawa 243-0435, Japan

‡Contributed equally to this work

## Contents

## 1. Data processing for fast Fourier transform (FFT) analysis of nano-FTIR line profiles

The raw profiles nano-FTIR line profiles contain single-pixel outliers caused by dirt particles on the α-quartz surface (left panels of Figure S1b, c). To remove these outliers, we divided the spectra into four frequency blocks (1096–1117 $cm^{-1}$, 1119–1140 $cm^{-1}$, 1142–1161 $cm^{-1}$, and 1163–1182 $cm^{-1}$) and averaged the profiles over frequency within each block. Figure S1a illustrates the procedure for the block of 1096–1117 $cm^{-1}$ block. Blue curve corresponds to the calculated average for each position in the profiles along the frequency axis. A threshold value (black line) was set to identify outliers. A pixel was classified as an outlier (black crosses) when the mean of its differences from the preceding and following pixels (orange and green curves in Figure S1a) exceeded this threshold. For each detected outlier position, we replaced the complex values defined by the amplitude and phase at each frequency with the average of the two neighboring pixels. The red dashed curve presents the averaged amplitude profile after correction. The correction results of all frequency blocks are shown in the right panels of Figure S1b, c. Although Figure S1 shows only the parallel configuration, the same procedure was applied to the perpendicular configuration.

After this correction process, we mirrored the real and imaginary parts of each line profile about the first pixel ($x = 0$) and concatenated them with the original data to suppress edge discontinuities in the FFT (Figure S2a). As shown in Figure S2b and S2c, we performed row-by-row baseline subtraction on the mirrored profiles. Figure S2b shows the baseline detected for each frequency row, and Figure S2c shows the baseline-subtracted profiles. We then applied the Tukey window function (Figure S3a) to the profiles in Figure S2c (Figure S3b) and zero-padded the data to obtain a finer wavevector sampling in the FFT along the $x$ axis. (Figure S3c). Finally, we performed the FFT to obtain the dispersion maps shown in Figure 4a of the main text. Figure S2 and S3 likewise show only the parallel configuration; the same procedure was applied to the perpendicular configuration.

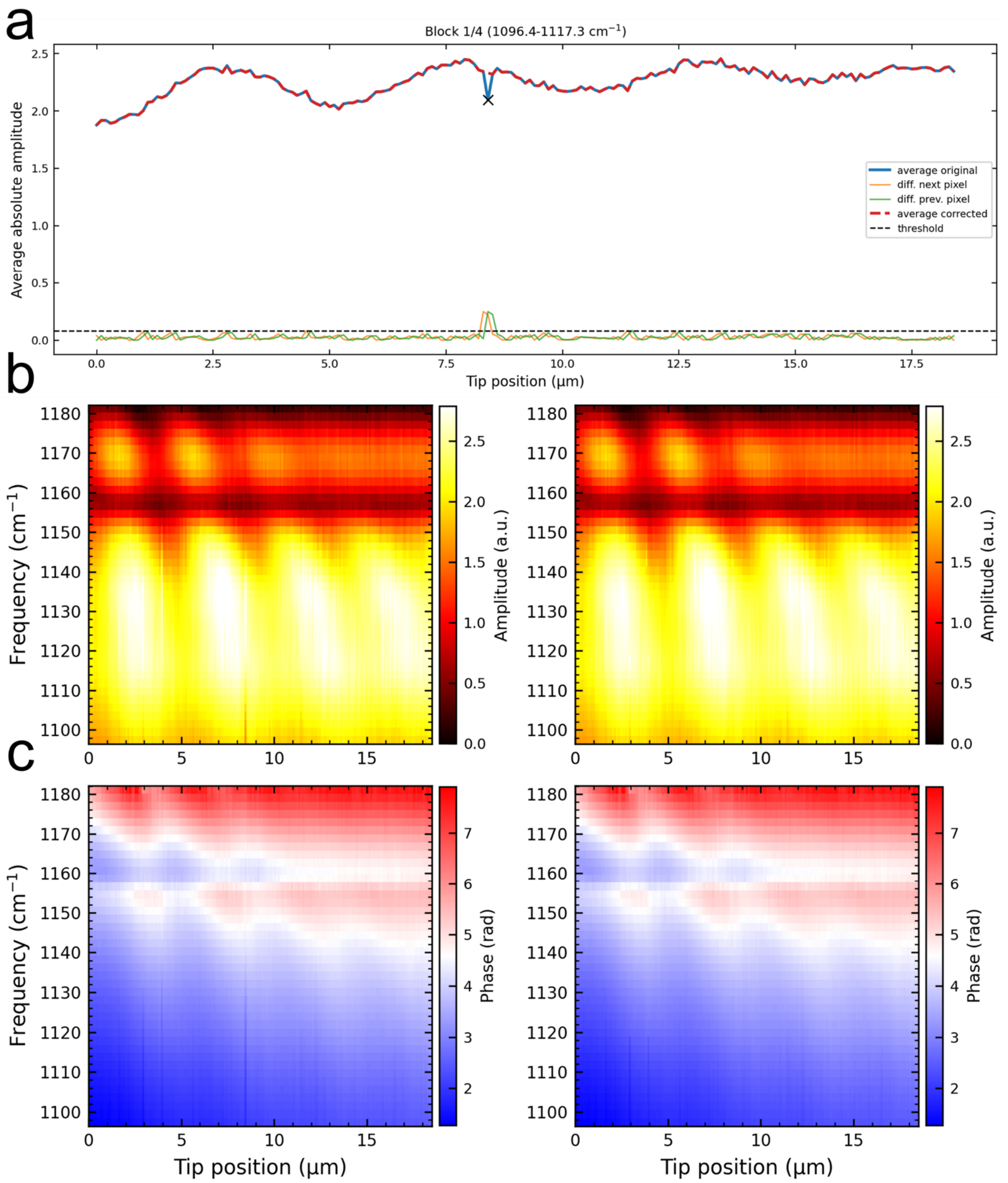

**Figure S6.** Outlier removal process. (a) Outlier detection and removal for the 1096–1117 $cm^{-1}$ block. Blue, orange, and green solid lines show the averaged profiles and its differences from the next and previous pixel, respectively. Black and red dashed lines show the threshold and averaged profile after correction, respectively. (b, c) Amplitude (b) and phase (c) profiles of SPhPs before removal (left) and after removal (right).

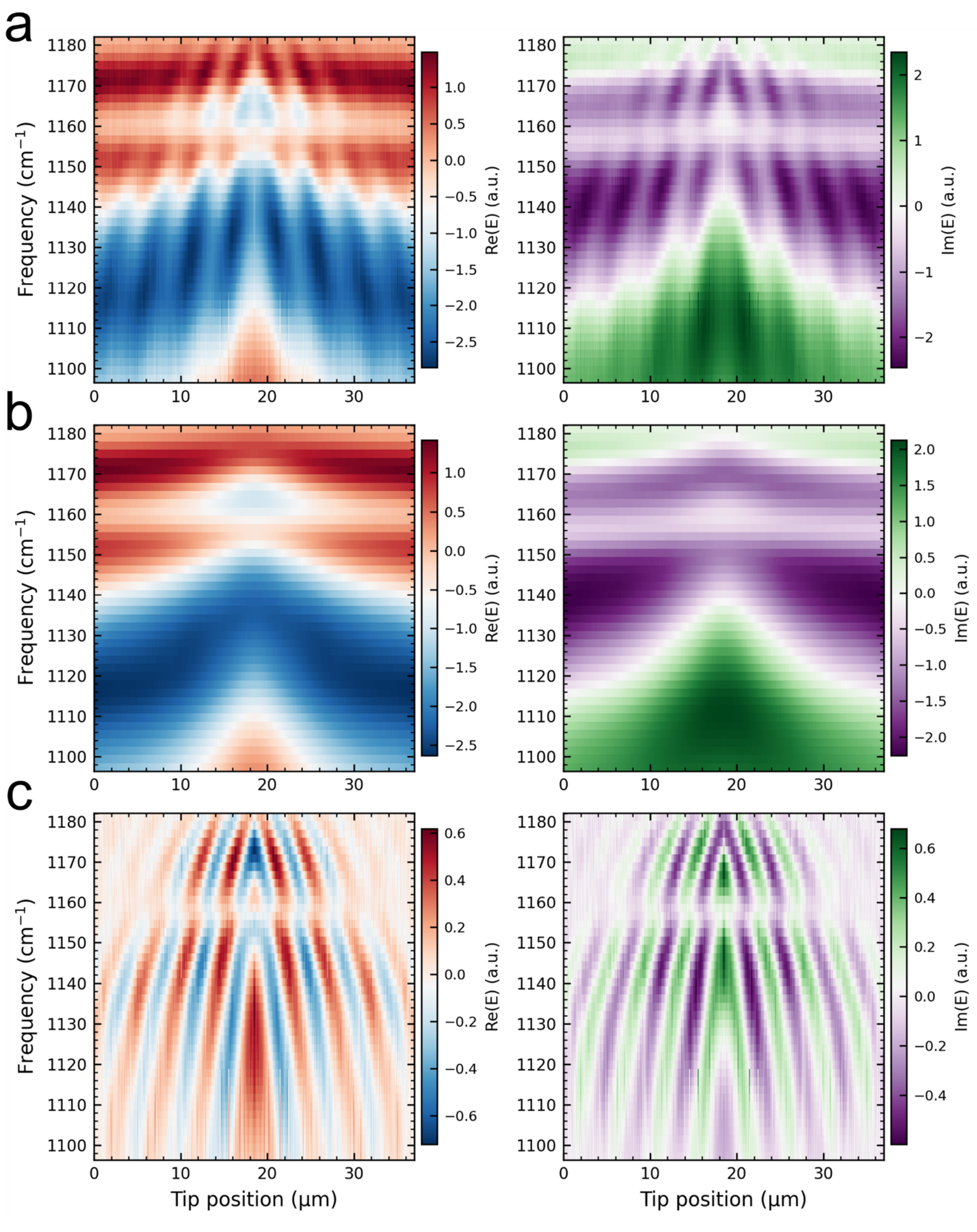

**Figure S7.** Mirroring and baseline subtraction process for the real (left) and imaginary (right) parts of the measured SPhP profiles. (a) Mirrored profiles. (b) Baseline detected in (a). (c) Baseline-subtracted profiles.

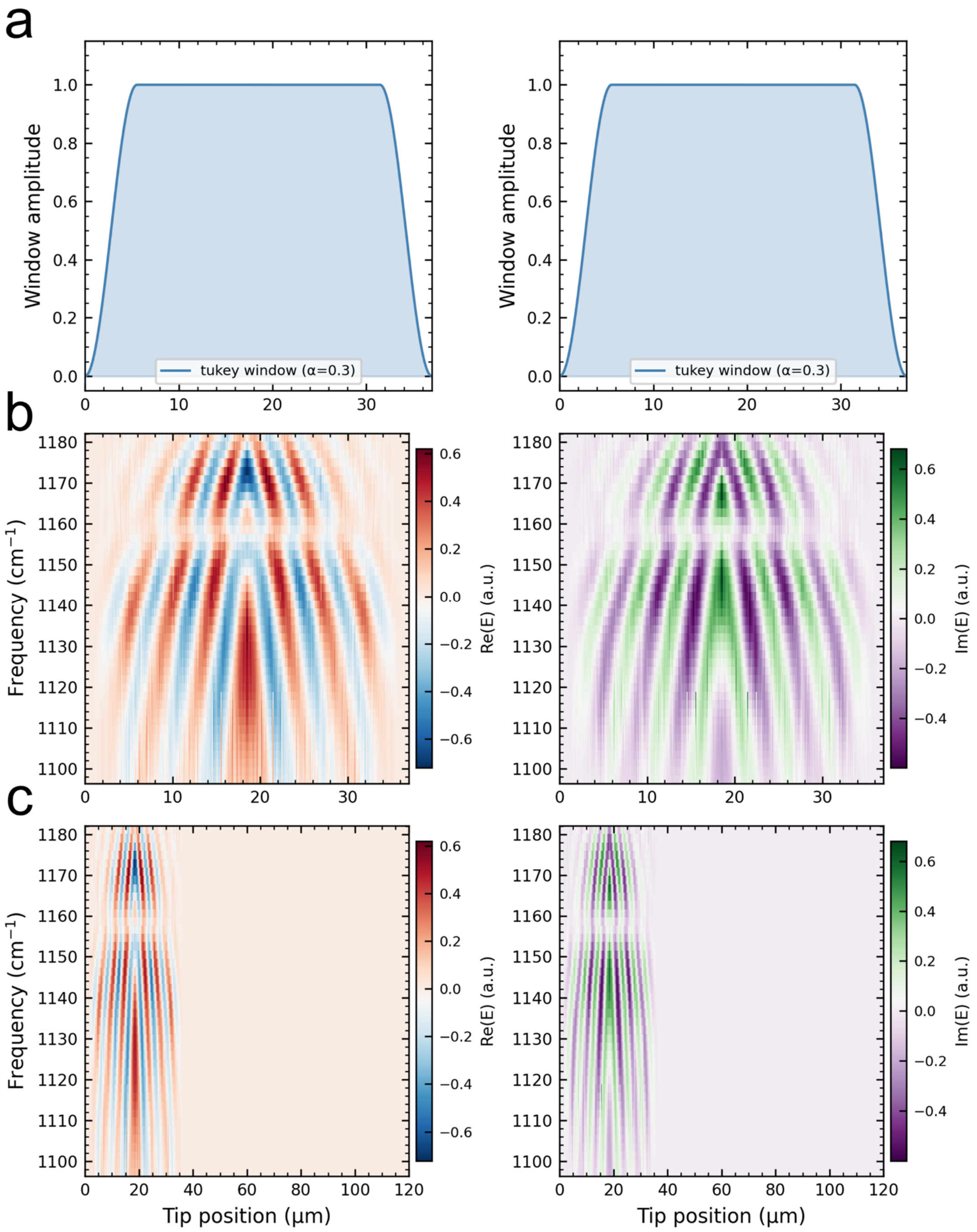

**Figure S8.** Window multiplication and zero padding process for measured SPhP profiles. (a) Tukey window function with $\alpha = 0.3$. (b) Tukey-windowed real (left) and imaginary (right) SPhP profiles. (c) Zero-padded real (left) and imaginary (right) SPhP profiles.

## 2. Reconstruction of edge-launched SPhPs using inverse FFT

To reconstruct the amplitude and phase profile maps of edge-launched SPhPs, we applied a window to the complex FFT data underlying Figure 4a of the main text. Figure S4a shows the Hann window centered on the calculated $k_{\mathrm{el}}$ curves in the parallel configuration (left) and the perpendicular configuration (right), respectively. The window half-width is 5000 rad/cm. The windowed data (Figure S4b) were then inverse FFT (iFFT). The reconstructed amplitude and phase maps obtained by the iFFT are shown in Figures 4b, c of the main text.

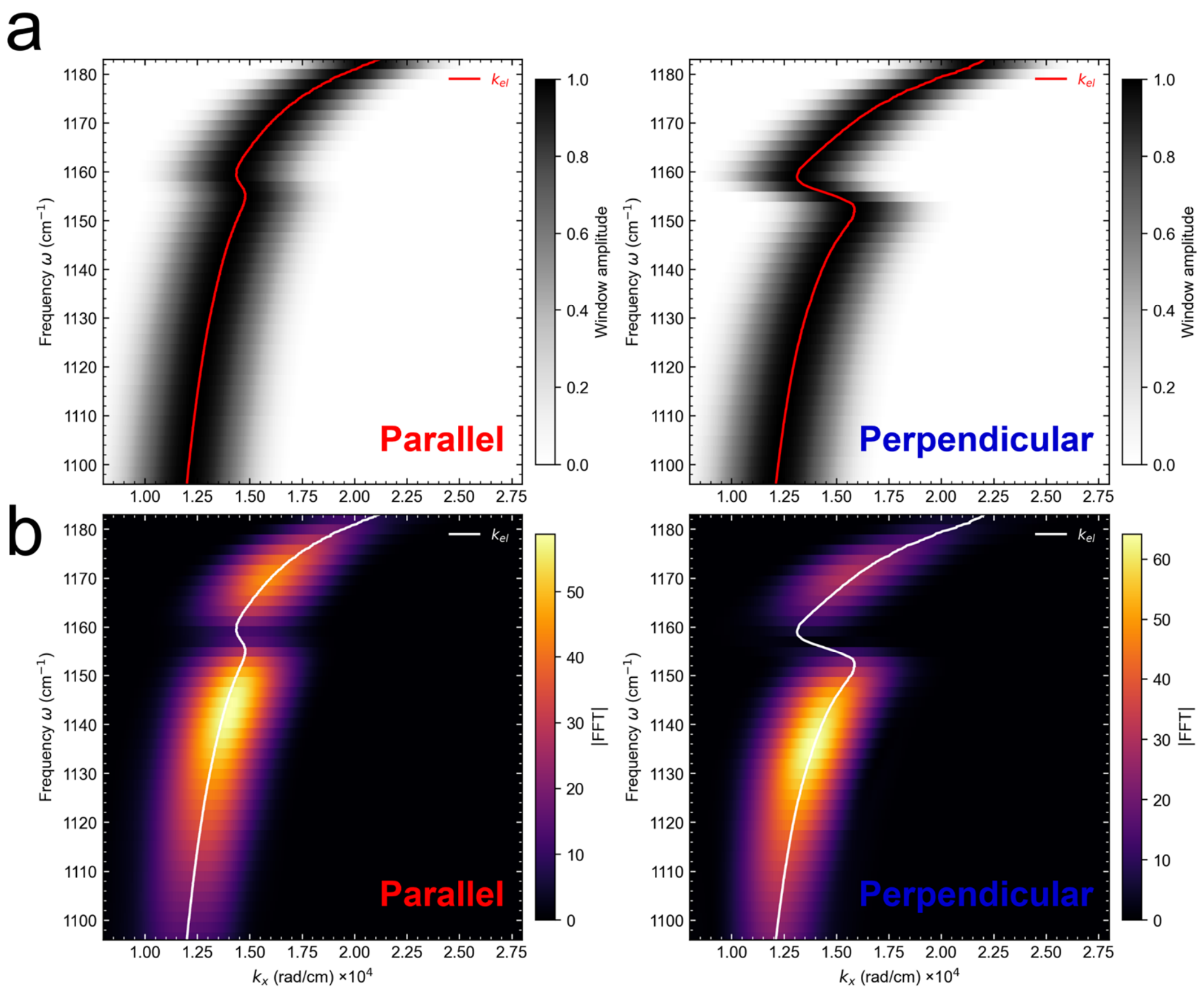

**Figure S9.** Hann-windowed FFT-magnitude maps for the parallel (left) and perpendicular (right) configurations. (a) Hann window centered the calculated $k_{\mathrm{el}}$ curves (red; half-width 5000 rad/cm). (b) Hann-windowed FFT-magnitude maps; white curves show the calculated $k_{\mathrm{el}}$. These maps were used to obtain iFFT profiles shown in Figures 4b, c of the main text.

## 3. Lorentzian fitting of step-windowed FFT-magnitude profiles

To determine the peak position $k_{\mathrm{el}}^{\mathrm{exp}}$ at each frequency, we fitted a Lorentzian function to the FFT-magnitude profiles. Figure S5a presents the step window function centered on the $k_{\mathrm{el}}$ dispersion curves for the parallel (left) and perpendicular (right) configuration. The window width is $\pm 5000$ rad/cm. To improve fitting precision, we multiplied the FFT-magnitude maps (Figure 4a of the main text) by the window function shown in Figure S5a. Figure S5b shows the step-windowed FFT-magnitude maps. Lorentzian fits to the windowed maps at each frequency yielded $k_{\mathrm{el}}^{\mathrm{exp}}$. The extracted $k_{\mathrm{el}}^{\mathrm{exp}}$ was then converted into $k_{\mathrm{SPhPs}}$ through the Eq. (7) of the main text. Representative Lorentzian fits at three frequencies are shown in Figure S6a (parallel) and S6b (perpendicular).

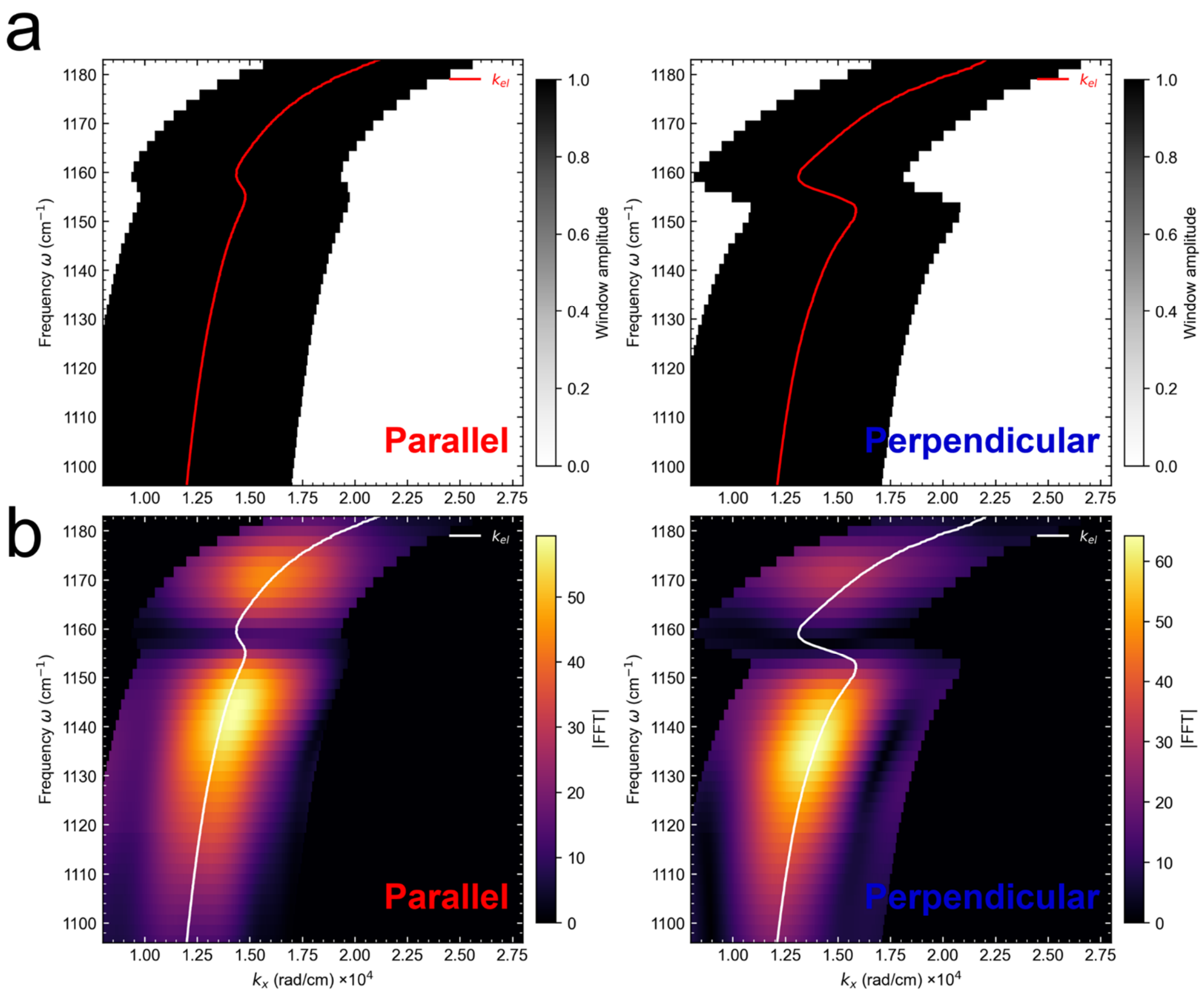

**Figure S10.** Step-windowed FFT-magnitude maps fot the parallel (left) and perpendicular (right) configuration. (a) Step window around the edge-launched SPhP rige (half-width 5000 rad/cm). The red curves show the calculated $k_{\mathrm{el}}$. (b) Step-windowed FFT-magnitude maps; white curves show the calculated $k_{\mathrm{el}}$. These maps were used to extract the $k_x$ peak positions $k_{\mathrm{el}}^{\mathrm{exp}}$.

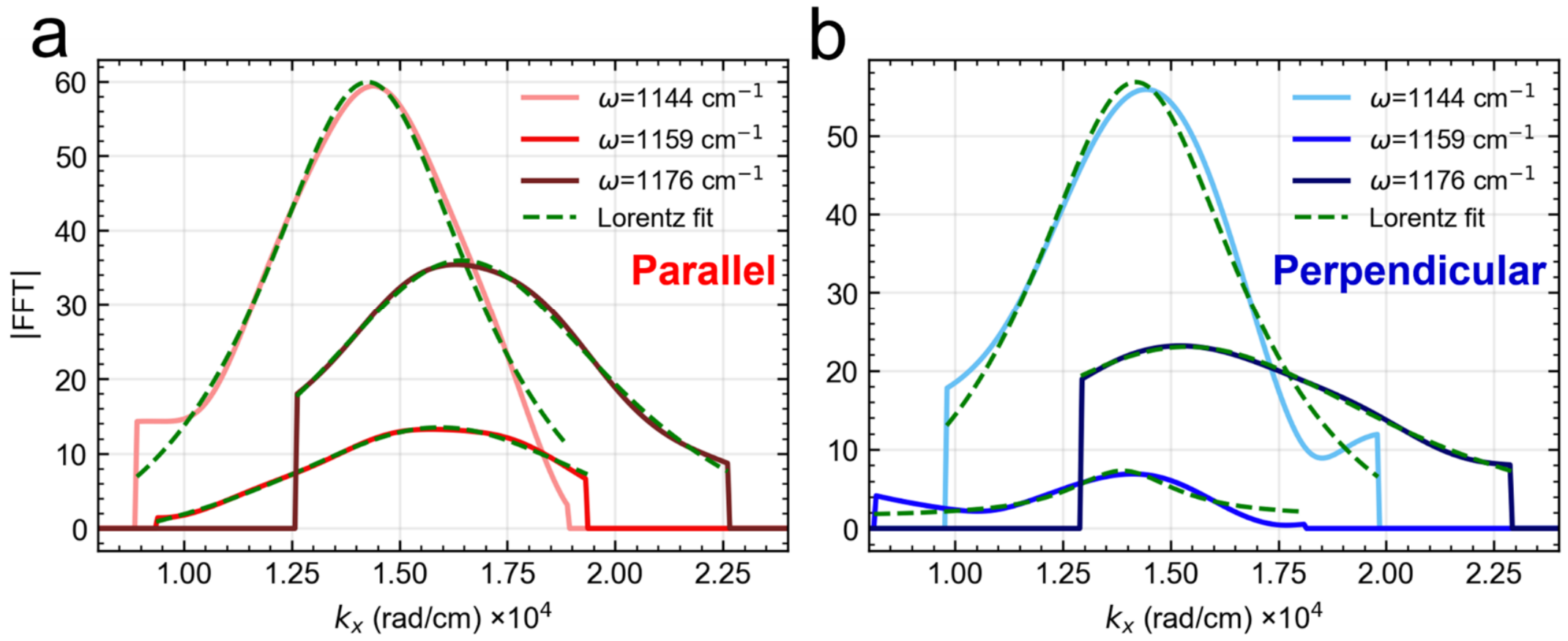

**Figure S11.** Example of the Lorentzian fits to the edge-launched SPhP peak at 1144, 1159, and 1176 cm$^{-1}$ for the (a) parallel and (b) perpendicular configuration.

## 4. Numerical simulations by finite-elements method

Finite-elements simulations (COMSOL Multiphysics®) were performed to calculate the complex-valued electric field of the edge-launched SPhPs. As shown in Figure S7a, an Au disk of 65 nm thickness and 10 μm diameter was placed at $z = 0$ on α-quartz and p-polarized light ($E_{\mathrm{in}}$) was incident at a polar angle 45° from the $+x$ direction. The principle components of the α-quartz permittivity tensor were described by the oscillator model of ref S1 (ref 31 of the main text):

$$\varepsilon(\omega) = \varepsilon_\infty + \sum_{n=1}^{n_{\max}} \frac{A_n \omega_n \Gamma_n}{\omega_n^2 - \omega^2 - i\omega\Gamma_n}, \tag{S1}$$

where $\varepsilon_\infty$ is the high-frequency permittivity, and $\omega_n$, $\Gamma_n$, and $A_n$ are the frequency, damping rate, and amplitude of the $n$th oscillator, respectively. These parameters are listed in Table S1.

Figure S7b and S7c present the electric field profiles of $\mathrm{Re}(E_z)$ and $\mathrm{Im}(E_z)$ between 1096~1183 $\mathrm{cm^{-1}}$ along $x$-axis at $z = 1$ nm. The disk edge is located at $x = 0$. Fourier transforming these complex profiles yielded the FFT-magnitude maps of $E_z$ shown in Figure S8. Because COMSOL adopts the $\exp(+i\omega x)$ time convention, a wave propagation in the $+x$ direction appears as $\exp(-ikx)$, opposite to the $\exp(ikx)$ convention used in our analysis. We therefore Fourier transformed the complex conjugate of the simulated $E_z$. The dashed white line denotes the light line $k_0 = \omega/c$. The SPhP wavevector extracted from the FFT-magnitude peaks are plotted as red and blue circles in Figure S8a and S8b, respectively, and as solid curves in Figure 5 of the main text.

**Table S1**. Oscillator model parameters of α-quartz determined by ellipsometry (ref S1).

| | Parallel to optic axis | | | | Perpendicular to optic axis | | |
|---|---|---|---|---|---|---|---|
| $n$ | $A_n$ | $\omega_n$ (cm$^{-1}$) | $\Gamma_n$ (cm$^{-1}$) | $n$ | $A_n$ | $\omega_n$ (cm$^{-1}$) | $\Gamma_n$ (cm$^{-1}$) |
| 1 | 68.024 | 363.93 | 3.6 | 1 | 44.021 | 393.61 | 3.3 |
| 2 | 0.212 | 415.64 | 15.5 | 2 | 101.700 | 449.84 | 3.7 |
| 3 | 82.130 | 494.17 | 4.5 | 3 | 1.906 | 694.89 | 8.3 |
| 4 | 12.639 | 776.03 | 7.2 | 4 | 12.722 | 794.88 | 7.5 |
| 5 | 120.713 | 1073.38 | 6.3 | 5 | 105.396 | 1065.23 | 6.9 |
| 6 | 0.040 | 1189.25 | 16.1 | 6 | 0.016 | 1142.33 | 3.6 |
| 7 | 0.027 | 1237.95 | 54.5 | 7 | 1.258 | 1160.85 | 8.0 |
| 8 | 0.052 | 1275.09 | 120.1 | 8 | 0.048 | 1175.36 | 176.8 |
| 9 | 0.042 | 1609.97 | 35.4 | 9 | 0.038 | 1270.83 | 39.4 |
| 10 | 0.026 | 1688.31 | 24.6 | 10 | 0.041 | 1875.00 | 34.7 |
| 11 | 0.012 | 1791.12 | 26.2 | | | | |
| 12 | 0.053 | 1876.05 | 43.6 | | | | |
| | $\varepsilon_\infty$ | 2.412 | | | $\varepsilon_\infty$ | 2.3839 | |

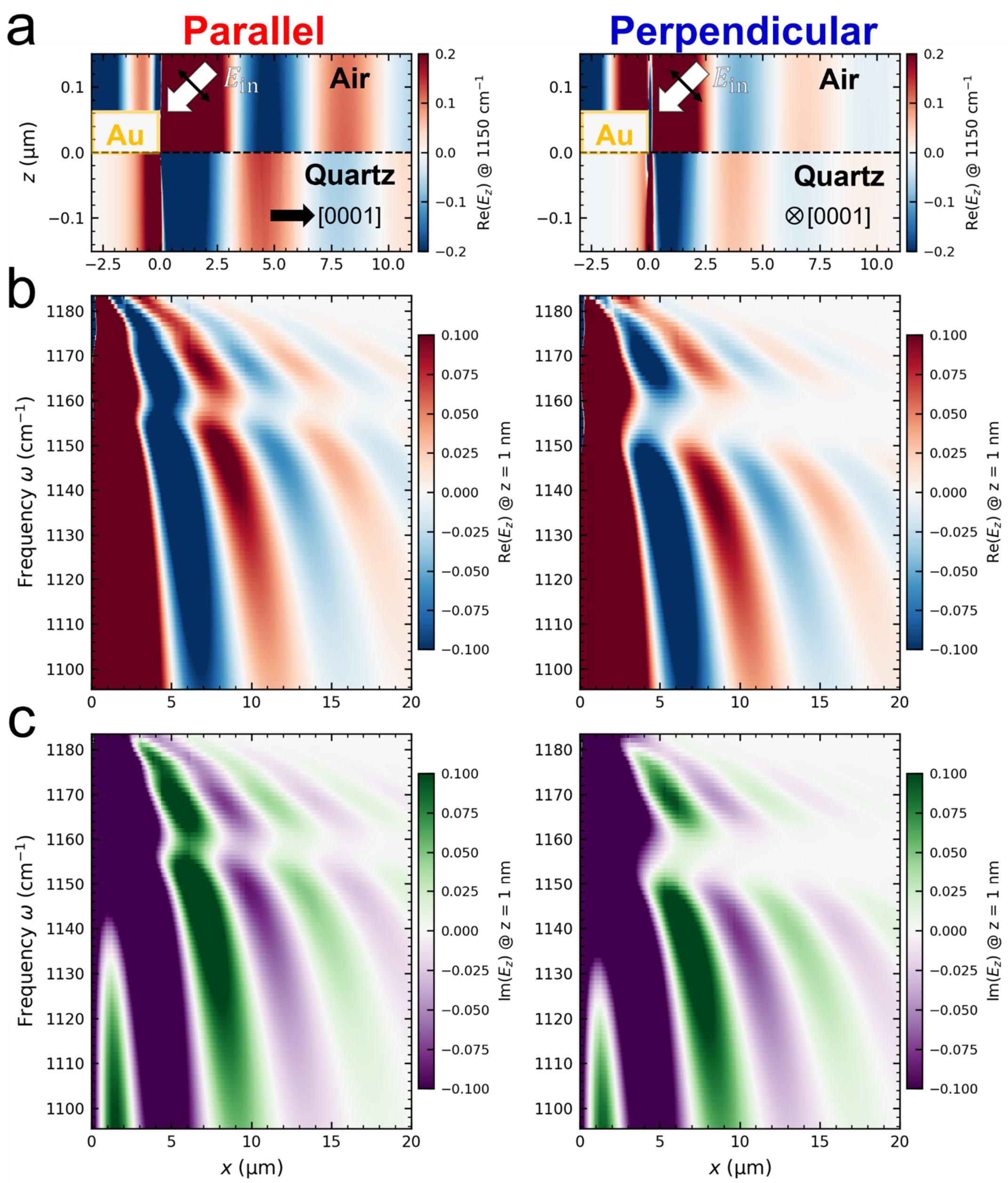

**Figure S12.** FEM simulations of edge-launched SPhPs on α-quartz for the parallel (left) and perpendicular (right) configurations. (a) $\mathrm{Re}(E_z)$ in the XZ-plane at $y = 0$ and $\omega = 1150$ cm$^{-1}$. (b) $\mathrm{Re}(E_z)$ and (c) $\mathrm{Im}(E_z)$ profiles along the $x$ axis at $z = 1$ nm from 1096 to 1183 cm$^{-1}$.

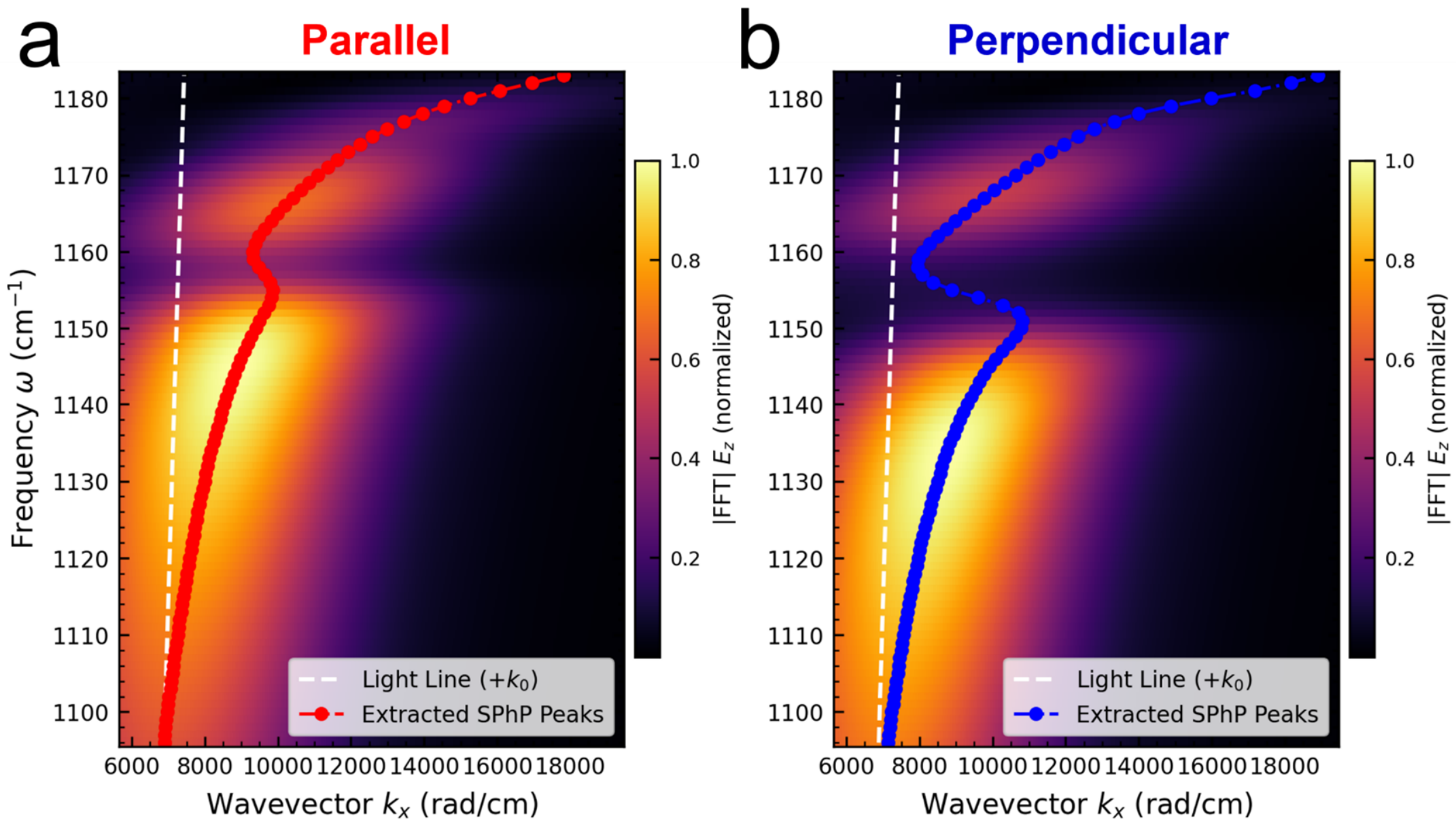

**Figure S13.** Simulated SPhP dispersion on α-quartz for the (a) parallel and (b) perpendicular configurations. Background maps show FFT-magnitude of $E_z$ obtained by complex FFT of the profiles in Figure S7b, c, along $x$-axis. The dashed white line denotes the light line $k = \omega/c$. Red and blue circles mark the FFT-magnitude peak positions at each frequency, which gives the SPhP dispersion for the parallel and the perpendicular configurations, respectively.